# Gallium Nitride Photodetector Measurements of UV Emission from a Gaseous CH4/O2 Hybrid Rocket Igniter Plume

Hannah S. Alpert
Stanford University
496 Lomita Mall
Stanford, CA 94305
halpert@stanford.edu

Ananth Saran Yalamarthy
Stanford University
496 Lomita Mall
Stanford, CA 94305
ananthy@stanford.edu

Peter F. Satterthwaite
Stanford University
496 Lomita Mall
Stanford, CA 94305
psatt99@stanford.edu

Elizabeth Jens
Jet Propulsion Laboratory,
California Institute of Technology
4800 Oak Grove Dr.
Pasadena, CA 91109
Elizabeth.Jens@jpl.nasa.gov

Jason Rabinovitch
Jet Propulsion Laboratory,
California Institute of Technology
4800 Oak Grove Dr.
Pasadena, CA 91109
Jason.Rabinovitch@jpl.nasa.gov

Noah Scandrette
Dept. of Physics and Astronomy
San Francisco State University
San Francisco, CA 94132
noahagape@gmail.com

AKM Newaz
Dept. of Physics and Astronomy
San Francisco State University
Thornton Hall, Room 317
1600 Holloway Avenue
San Francisco, CA 94132
akmnewaz@sfsu.edu

Ashley C. Karp
Jet Propulsion Laboratory,
California Institute of Technology
4800 Oak Grove Dr.
Pasadena, CA 91109
Ashley.C.Karp@jpl.nasa.gov

Debbie G. Senesky
Stanford University
496 Lomita Mall
Stanford, CA 94305
dsenesky@stanford.edu

*Abstract*—Owing to its wide (3.4 eV) and direct-tunable band gap, gallium nitride (GaN) is an excellent material platform to make UV photodetectors. GaN is also stable in radiation-rich and high-temperature environments, which makes photodetectors fabricated using this material useful for in-situ flame detection and combustion monitoring. In this paper, we use a GaN photodetector to measure ultraviolet (UV) emissions from a hybrid rocket motor igniter plume. The GaN photodetector, built at the Stanford Nanofabrication Facility, has 5 µm wide regions of AlGaN/GaN two-dimensional electron gas (2DEG) electrodes spaced by intrinsic GaN channels. In most applications, the ideal photodetector would exhibit a high responsivity to maximize the signal, in addition to a low dark current to minimize quiescent power. A performance metric which simultaneously captures these two values is the normalized photocurrent-to-dark current ratio (NPDR), defined as the ratio of responsivity to dark current, with units of $W^{-1}$. The NPDR of our device is record-high with a value of $6 \times 10^{14}$ $W^{-1}$ and the UV-to-visible rejection ratio is $4 \times 10^6$. The high rejection ratio is essential as it eliminates cross-sensitivity of the detector to visible light. The spectral response can be modeled as a rectangular window with a peak responsivity of 7,800 $AW^{-1}$ at 362 nm and a bandwidth of 16 nm. The photodetector shows operation at high temperatures (up to 250°C). The NPDR still remains above $10^9$ $W^{-1}$ at the higher temperatures, and the peak wavelength shifts from 362 nm to 375 nm at 250°C. The photodetector was placed at three radial distances (3", 5.5", and 7") from the base of the igniter plume and the oxidizer-to-fuel ratio ($O_2/CH_4$) was varied to alter the size and strength of the plume. The current measured from the device was proportional to the intensity of the emission from the plume. The data demonstrates a clear trend of increasing current with increasing fuel concentration. Further, the current decreases with larger separation between the photodetector and the plume. A calibration curve constructed from the responsivity measurements taken over four orders of magnitude was used to convert the current into incident optical power. By treating the plume as a black body, and calculating a radiative configuration factor corresponding to the geometry of the plume and the detector, we calculated average plume temperatures at each of the three oxidizer-to-fuel ratios. The estimated plume temperatures were between 850 and 950 K for all three combustion conditions. The temperature is roughly invariant for a fixed fuel concentration for the three tested distances. These data demonstrate the functionality of GaN as a material platform for use in harsh environment flame monitoring.

## Table of Contents

## 1. INTRODUCTION

Ultraviolet (UV) photodetectors have many diverse uses, such as chemical analysis for environmental applications, communication between satellites, UV astronomy, flame detection for fire alarms, and combustion monitoring [1],[2]. The parameters that determine photodetector performance include signal-to-noise ratio, response time, dark current, and responsivity [2], [3]. Responsivity is the ratio of the photocurrent to the incident optical power. The responsivity divided by the dark current is known as the normalized photocurrent-to-dark current ratio (NPDR), which incorporates the parameters that govern a large signal amplitude and low quiescent power in a single performance metric [4]-[6]. Another important metric is the UV-to-visible rejection ratio, which describes the cross-sensitivity of the detector to visible light.

Optical measurement of flame temperature is typically done using two photodetectors whose spectral responsivities peak at different wavelengths [7]. The ratio of the emission intensities measured by the two photodetectors changes with flame temperature, with the key advantage being that this measurement is not affected by the area of the emitting source. Flame temperature sensors in literature accomplish this dual-spectrum response using a variety of strategies: filtering one half of the photodetector to block certain wavelengths while leaving the other half unfiltered [8],[9], varying the alloy content in photodetectors made with heterostructure materials to change the absorptivity and wavelength of peak responsivity [10], or using two separate photodetectors altogether that have different optical properties [11].

Gallium nitride (GaN) is an excellent material platform to make UV photodetectors due to its wide (3.4 eV) and direct-tunable band gap [10] and high responsivity [3]. GaN is also more stable than frequently used materials (e.g., silicon) in radiation-rich [3],[12],[13] and high-temperature [10],[12],[14] environments. This makes photodetectors fabricated with GaN uniquely useful for in-situ flame detection and combustion monitoring.

In this paper, we use a GaN photodetector with a record-high NPDR of $6 \times 10^{14}$ $W^{-1}$, a peak responsivity of 7,800 A/W, and a UV-to-visible rejection ratio of $4 \times 10^6$, to measure UV emissions from a hybrid rocket motor igniter plume and thus estimate its temperature. In these experiments, the photodetector is at room temperature; however, we have also separately characterized the optical response up to high temperatures (250°C). In addition to high responsivity, the small size, high-temperature operation, and extremely low quiescent power consumption (~10s of pW) of these photodetectors make them good candidates for use in the proximity of flames and combustion chambers, as well as in other harsh environments, such as entry descent and landing systems.

We have organized this paper as follows: Section 2 outlines the fabrication and operation of the photodetectors, Section 3 describes high temperature operation of the photodetectors, Section 4 discusses the experimental setup, section 5 contains a discussion of the results, and section 6 provides a summary of the work.

## 2. DEVICE FABRICATION AND OPERATION

*Fabrication*

The microfabrication process for the GaN photodetectors was previously reported by us elsewhere [6], and a brief description is provided here for clarity. The devices were fabricated on aluminum gallium nitride (AlGaN)/GaN-on-Si wafers that were grown by metal organic chemical vapor deposition (MOCVD); the III-nitride stack is depicted in Figure 1b. The AlGaN/GaN interface is host to a two-dimensional electron gas (2DEG) with high channel mobility (~2000 $cm^2V^{-1}s^{-1}$ at room temperature). A reactive ion etch was used to form an array of 2DEG interdigitated transducers (IDT). The IDT is comprised of AlGaN fingers separated by GaN channels. The AlGaN/GaN finger electrodes are contacted by annealed Ti/Al/Pt/Au Ohmic contacts. The detailed operation of the device is discussed elsewhere [6]. The final device, with an active area of 280 µm x 200 µm, has alternating 5 µm wide regions of 2DEG electrodes and intrinsic GaN channels, as shown in Figure 1a.

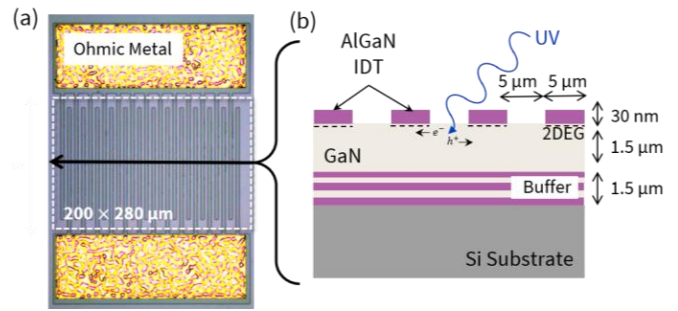

**Figure 1. (a) 3D cross-section of the UV photodetector. (b) Optical microscope image of the device.**

*Device Characterization*

In order to understand the optical properties of the device, we characterized its spectral response across a wide range of incident powers, from ~$10^4$ mW/cm$^2$ to ~1.5 mW/cm$^2$. An example of the spectral response at an optical power of ~$4 \times 10^{-3}$ mW/cm$^2$ is illustrated in Figure 2. Here, it is seen that the responsivity peaks at a wavelength of ~362 nm. Further, the spectral response can be modeled as a rectangular window with a height corresponding to the peak responsivity at 362 nm (433 A/W) and a width of 16 nm. Figure 2 also shows a UV-to-visible (362 nm to 425 nm) ratio of 4 x 10⁶. The variation in the responsivity across the range of incident optical powers at a wavelength of 362 nm is illustrated in Figure 3. The peak responsivity of ~7800 A/W occurs at a power of ~1 mW/cm². Since the area of the device is known, we can convert the responsivity into a measurement of the current across the device, which yields a calibration curve of the current versus the total optical power on the device, as shown in Figure 4. This calibration curve can be used to extract the flame temperature from the incident optical power on the photodetector.

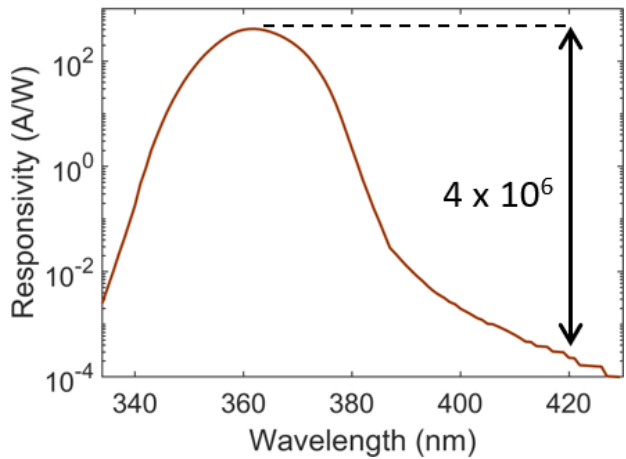

**Figure 2. Spectral responsivity vs. wavelength, measured at a bias voltage of 5 V and incident power of $10^5$ mW.**

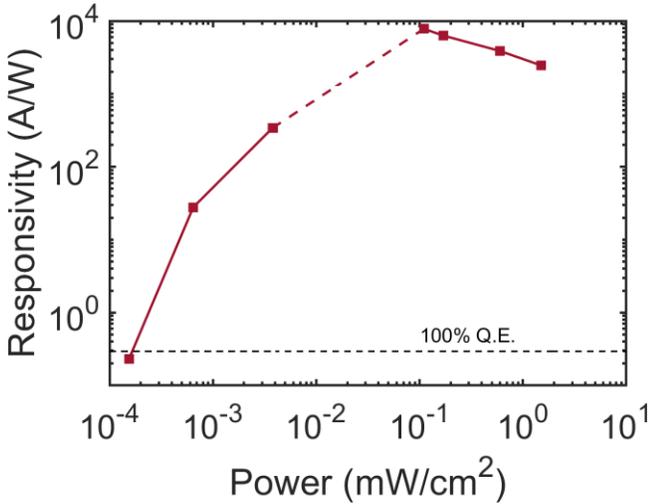

**Figure 3. Plot of responsivity versus incident power for a wide range of incident optical powers.**

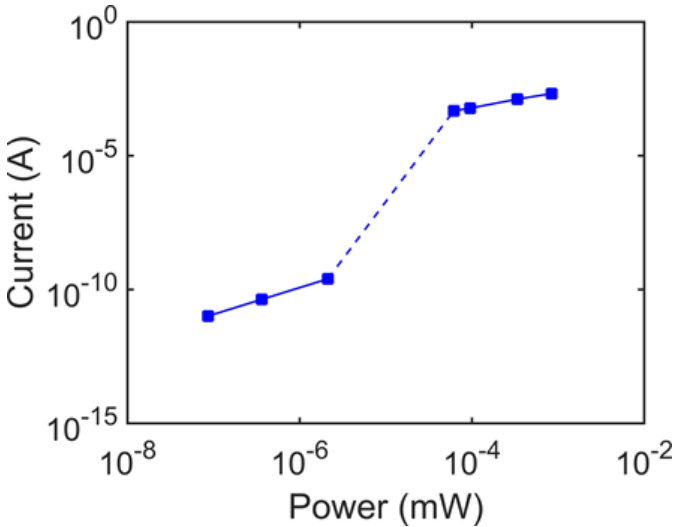

**Figure 4. Measurements of current versus incident power for a wide range of incident optical powers.**

## 3. HIGH TEMPERATURE OPERATION

The high temperatures properties of the GaN photodetector were investigated to evaluate its performance in extreme environment conditions. In these experiments, we fixed the incident optical power at ~4 x 10⁻³ mW/cm², although a similar trend is expected at other optical powers. The peak spectral responsivity of the photodetector fell by a factor of nearly 2,000 from room temperature to 250°C (Figure 5a), but the NPDR still remained above $10^9$ W⁻¹ even at the highest temperatures (Figure 5b), which is still higher than the values of several GaN photodetectors at room temperature in the published literature [15],[16]. In particular, this NPDR of $10^9$ W⁻¹ at 250°C corresponds to a photocurrent-to-dark current ratio of ~20, and thus amenable to photodetection at these high temperatures. The steep fall in responsivity as the temperature increases is the result of a combination of lower mobility [17] of the charge carriers due to lattice scattering and a higher recombination rate for the photo-excited electron-hole pairs. The wavelength at which the highest responsivity occurs shifted from ~362 nm at room temperature to ~375 nm at 250°C, as shown in Figure 6a. This wavelength shift is due to the decrease in bandgap with increasing temperature. For GaN, the bandgap, in units of eV, is a function of temperature ($T$), and can be expressed using the Varshni parameters as [18]:

$$E_g = 3.51 - \frac{9.09 * 10^{-4} * T^2}{T + 830}. \qquad (1)$$

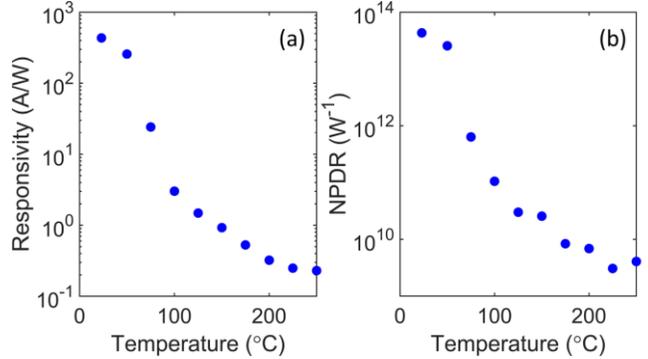

**Figure 5. Plots of (a) responsivity and (b) Normalized photocurrent to dark current ratio (NPDR) versus temperature.**

The corresponding wavelengths for these temperature-dependent bandgaps are plotted in Figure 6b, showing excellent agreement with the experimentally measured wavelength where the peak responsivity occurs (Figure 6a). This confirms that the shift to higher wavelengths of the peak spectral responsivity with temperature is due to the change in bandgap.

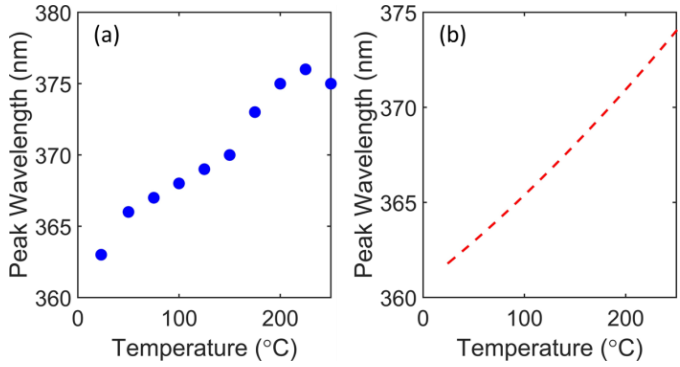

**Figure 6.** Plots of peak wavelength versus temperature for (a) measured data and (b) theoretical calculations.

## 4. EXPERIMENTAL METHODS

The GaN photodetector was placed at three radial distances (3", 5.5", and 7") from the base of a hybrid rocket motor igniter plume, and the $O_2/CH_4$ oxidizer-to-fuel (O/F) ratio was varied to alter the size and strength of the plume. An image of the test setup is shown in Figure 7a. UV light from the entire plume (shown in Figure 7b) is collected by the photodetector, since it is not collimated in our experimental setup. A more detailed description of the hybrid rocket motor and its characteristics can be found elsewhere [19]. A commercial spark plug was used to ignite the $O_2/CH_4$ mixture in the igniter combustion chamber. The feed pressure of oxygen was kept constant to provide a consistent mass flow rate of 0.70 g/s, while the methane pressure varied between 0.12 and 0.24 g/s to establish a range of O/F ratios. The igniter was fired under three conditions: lean, near stoichiometric, and rich, which correspond to O/F ratios by mass of 5.8, 4.5, and 2.9 respectively. The stoichiometric O/F ratio for the $O_2/CH_4$ reaction is 4.0. The average chamber pressures and maximum chamber pressures also varied for the different conditions, detailed in Table 1. At every condition, the igniter was fired two to four times, with each burn lasting one second followed by a three second purge and a wait time of at least 30 seconds before the next burn. A sourcemeter (Keithley 2400) was used to bias the photodetector at 5 V and measure the device current at a sample rate of ~13.5 Hz. An image of the setup as well as a schematic of the photodetector and flame is shown in Figure 7. Each fire of the igniter caused a spike in the current proportional to the intensity of the UV emission from the flame, and an example of this response is shown in Figure 8. We note that in these experiments, the photodetector is at room temperature due to the short duration of the fire and the large spatial separation from the plume.

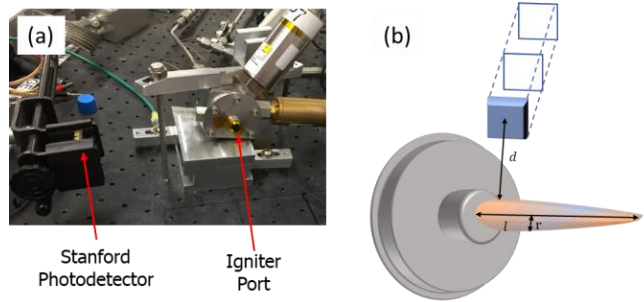

**Figure 7.** (a) Image of photodetector setup. The sensor is being held by the clamp on the left and the igniter plume is emitted from the hole in the middle. (b) Schematic of flame and photodetector where *d* is the radial distance between the two, *l* is the length of the flame, and *r* is the radius of the flame.

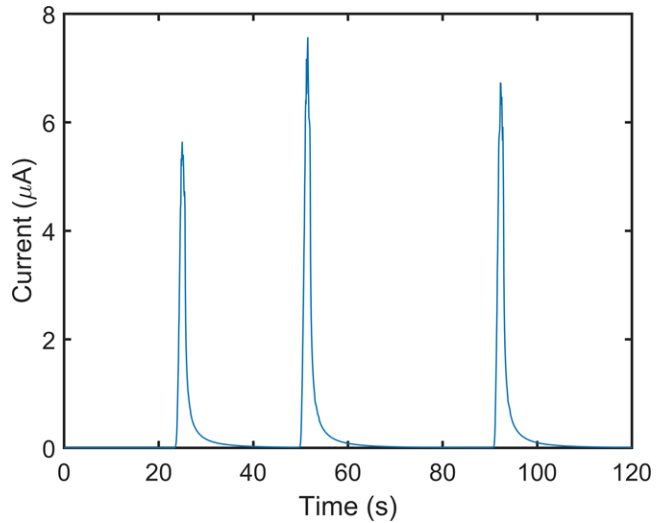

**Figure 8.** Plot of current measured across the photodetector versus time for a series of three fires.

|  | **Lean** | **Near Stoichiometric** | **Rich** |
|---|---|---|---|
| CH4 Mass Flow Rate (g/s) | 0.12 | 0.15-0.16 | 0.24 |
| Oxidizer-to-Fuel Ratio | 5.81-5.96 | 4.47-4.62 | 2.85-2.89 |
| Average Chamber Pressure (psi) | 70.0-72.8 | 82.6-84.4 | 104.6-109.1 |
| Maximum Chamber Pressure (psi) | 89.7-93.3 | 105.4-107.5 | 133.3-138.1 |

**Table 1.** Values associated with various test conditions.

## 5. RESULTS AND DISCUSSION

For the matrix of O/F ratios and distances of the photodetector from the plume, we converted the current into incident optical power using the calibration data in Figure 4. The data from the photodetector demonstrates a clear trend of increasing optical power with increasing fuel concentration, as seen in Figure 9. Further, the optical power decreases with larger separation between the photodetector and plume, since a smaller fraction of the UV radiation is incident on the photodetector.

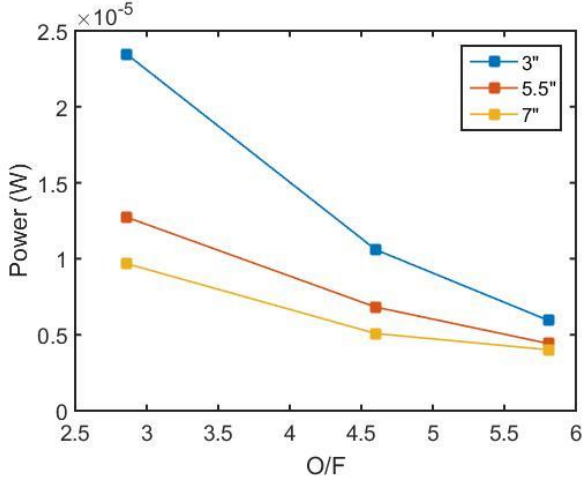

**Figure 9. Plot of incident optical power versus O/F ratio, showing a trend of increasing optical power with increasing fuel concentration, and decreasing optical power with larger separation between photodetector and plume.**

*Calculation of Flame Temperature*

The relationship between temperature and blackbody emissive power was used to estimate an average flame temperature, and a radiative configuration factor corresponding to the geometry of the emitter and the detector was incorporated to improve the model. In the model for radiative heat transfer between the plume and the photodetector, we assume that the plume and the photodetector can be approximated by a cylinder and rectangle (of size 200 µm by 280 µm), respectively. The normal to the rectangle passes through one end of the cylinder and is perpendicular to its axis, with dimensions as shown in Figure 7b.

The intensity (*i*) of light at a given wavelength ($\lambda$) is a function of the temperature (*T*) of the emitter. It is defined as the radiation emitted by a blackbody at temperature *T* per unit time, per unit surface area, per unit wavelength, with units of W/m²·µm

$$i(\lambda, T) = \frac{2C_1}{\lambda^5 \left(e^{\frac{C_2}{\lambda T}} - 1\right)}. \tag{2}$$

The constants $C_1$ and $C_2$ are defined as $hc^2$ and $hc/k_B$ respectively, where *h* is Planck's constant, *c* is the speed of light, and $k_B$ is the Boltzmann constant. The total intensity from a specific band of the electromagnetic spectrum can be calculated by integrating over the wavelengths of interest. In this case, the responsivity of the photodetector is a rectangular window centered around 362 nm (since it is at room temperature) with a width of 16 nm. Thus the power (*P*) incident on the photodetector, accounting for its spectral response, is given by:

$$P = \pi * i(T) * A_1 \tag{3}$$

where $A_1$ is the surface area of the emitter. A configuration factor is introduced to account for the effects of the geometry and orientation of the emitter and the detector. The configuration factor $F_{ij}$ is the fraction of the radiation leaving the surface of the emitter (with surface area $A_i$) and striking the surface of the detector (with area $A_j$). The reciprocity relation for configuration factors states that $A_i F_{ij} = A_j F_{ji}$ [20]. The configuration factor for a square emitter and cylindrical detector is given in [21]; the reciprocity relation must then be used to attain the configuration factor for the cylindrical emitter (with surface area $A_1$) and square detector (with area $A_2$), resulting in the following:

$$F_{12} = \frac{A_2}{A_1} * \left[ \frac{1}{\pi H} \tan^{-1} \frac{L}{\sqrt{H^2 - 1}} \right.$$
$$+ \frac{L}{\pi} \left( \frac{X - 2H}{H\sqrt{XY}} \right) \tan^{-1} \sqrt{\frac{X(H-1)}{Y(H+1)}}$$
$$\left. - \frac{1}{H} \tan^{-1} \sqrt{\frac{H-1}{H+1}} \right] \tag{4}$$

where *L*, *H*, *X*, and *Y* depend on *l*, *r*, and *d* from Figure 7b.

$$L = \frac{l}{r} \tag{5}$$

$$H = \frac{d}{r} \tag{6}$$

$$X = (1 + H)^2 + L^2 \tag{7}$$

$$Y = (1 - H)^2 + L^2 \tag{8}$$

In this case, the area of the detector $A_2$ is 280 µm x 200 µm, and the surface area of the flame ($2\pi r l$) varied with O/F ratio, increasing as the combustion reaction increased in fuel content. Multiplying (3) by the configuration factor $F_{12}$ results in the power emitted by the plume and detected by the photodetector.

The estimated plume temperatures contain considerable uncertainty. The flame area was challenging to quantify accurately, and in the lean condition, the flame was barely visible. Additionally, the surface area of the flame based on visible light may be different than that for UV light. To eliminate the uncertainty of the flame area, the light needs to be collimated in future experiments to provide a known area. Further, the model assumes that the temperature of the flame is uniform, but the temperature actually varies significantly throughout the plume [22]. In spite of these issues, the temperature was roughly invariant (within 25 K) for each fixed fuel concentration for the three tested distances (Figure 10). The theoretical nozzle temperatures for the test conditions are shown in Figure 11, with the two chamber pressures corresponding to the maximum and minimum seen in the experimental tests. The nozzle throat temperature varies by less than 100 K between the three O/F ratios, which was also seen in the calculated temperatures (inset of Figure 10). As seen in Figure 11, the nozzle temperatures are significantly higher than the calculated temperatures (from the photodetector). This is because the flame cools as it expands into the atmosphere, and the calculated temperatures are averages over the entire plume. Additionally, there is significant heat loss to the stainless steel igniter chamber itself, further reducing the temperature below the adiabatic flame temperature.

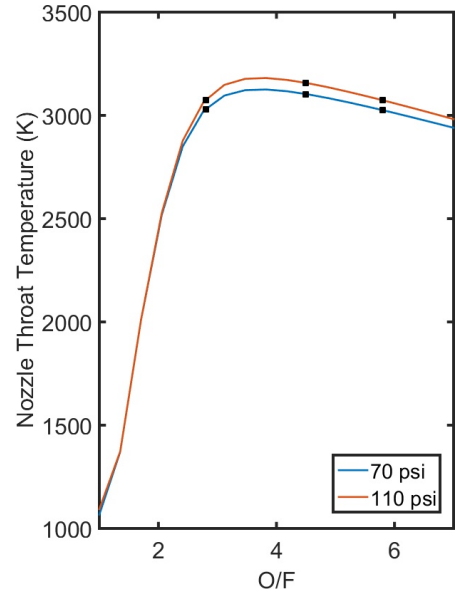

**Plot of theoretical nozzle temperatures over a range of O/F ratios for two different chamber pressures calculated using CEA.**

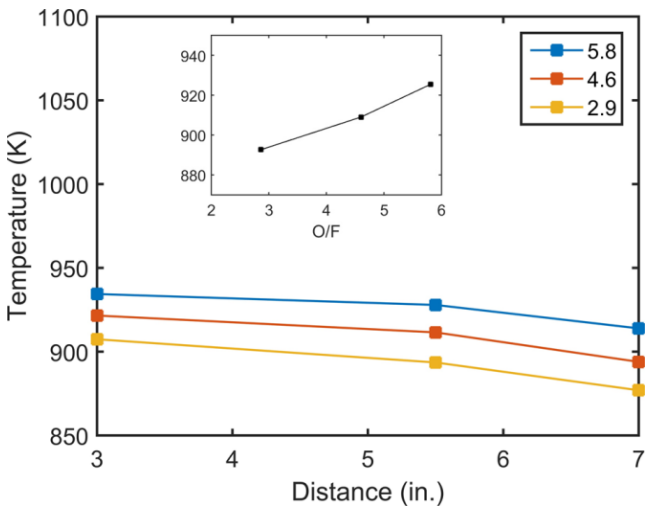

**Figure 10. Plot of calculated average plume temperature vs. distance of the photodetector from the plume for the three O/F ratios. The inset shows that the average calculated plume temperature for each O/F ratio was between 880 K and 940 K.**

## 6. CONCLUSIONS

We have shown the functionality of GaN photodetectors for use in flame detection and combustion monitoring of a hybrid rocket motor igniter plume. The sensors' small size, high-temperature operation, and low power consumption make them a good candidate for use near flames or combustion chambers, as well as in other harsh environments like entry descent and landing systems. While using a single photodetector to estimate flame temperature is difficult, GaN's direct-tunable bandgap suggests that it is a promising platform for a micro-scale temperature sensor by designing photodetectors with heterostructures in the III-nitride family. Further work will correlate plume conditions to combustion chamber activity and temperature, which may be used for in-situ monitoring of hybrid rocket motors and informing the design of future hybrid combustion chambers.

## ACKNOWLEDGEMENTS

The Jet Propulsion Laboratory, California Institute of Technology, provided funding and testing for this research under a contract with the National Aeronautics and Space Administration. This work was also funded by the National Science Foundation (Grant No. EECS-1708907). The photodetectors were fabricated at the Stanford Nanofabrication Facility (SNF).